\documentclass[12pt,english,sort&compress]{article}
\usepackage[T1]{fontenc}
\usepackage{textcomp}
\usepackage[latin9]{inputenc}
\usepackage{fancyhdr}
\usepackage{color}
\usepackage{comment}
\usepackage{babel}
\usepackage{array}
\usepackage{booktabs}
\usepackage{varwidth}
\usepackage{amsmath}
\usepackage{graphicx}
\usepackage[a4paper]{geometry}
\usepackage{setspace}
\usepackage[numbers]{natbib}
\usepackage[pdfusetitle,
 bookmarks=true,bookmarksnumbered=false,bookmarksopen=false,
 breaklinks=false,pdfborder={0 0 0},pdfborderstyle={},backref=false,colorlinks=false]
 {hyperref}

\makeatletter

\providecommand{\tabularnewline}{\\}
\newenvironment{cellvarwidth}[1][t]
    {\begin{varwidth}[#1]{\linewidth}}
    {\@finalstrut\@arstrutbox\end{varwidth}}

\usepackage{times}

\date{}
\renewcommand{\@openbib@code}{\setlength{\itemsep}{-1pt}}

\renewcommand{\subsectionmark}[1]{}

\usepackage[compact]{titlesec}
\titleformat{\section}{\LARGE \bfseries}{\thesection}{1em}{}
\titleformat{\subsection}{\large \bfseries}{\thesubsection}{1em}{}
\makeatother

\begin{document}
\global\long\def\d{\mathrm{d}}%
\global\long\def\id{\mathbf{I}}%

\global\long\def\defgradT{\mathbf{F}}%
\global\long\def\xh{\hat{\mathbf{x}}}%
 
\global\long\def\yh{\hat{\mathbf{y}}}%
 
\global\long\def\zh{\hat{\mathbf{z}}}%

\global\long\def\lueref{L}%
\global\long\def\cref{c_{0}}%

\global\long\def\macchainref{R}%
\global\long\def\links{n}%
\global\long\def\bonds{m}%
 
\global\long\def\maxbonds{m_{0}}%
 
\global\long\def\chainref{R_{f}}%
\global\long\def\linklen{l_{s}}%

\global\long\def\luecur{l}%
\global\long\def\bcur{\delta_{b}}%
\global\long\def\ccur{c}%
\global\long\def\macchaincur{r}%
\global\long\def\chaincur{r_{f}}%

\global\long\def\sT{\mathbf{S}}%
\global\long\def\rotationT{\mathbf{Q}}%

\global\long\def\dirTref{\hat{\mathbf{L}}}%
\global\long\def\dirTcur{\hat{\mathbf{l}}}%
\global\long\def\rTvec{\hat{\mathbf{l}}_{0}}%

\global\long\def\stretch{\lambda}%
\global\long\def\stretchue{\lambda_{re}}%
\global\long\def\ang{\theta}%

\global\long\def\funcone{g_{1}}%
\global\long\def\functwo{g_{2}}%

\global\long\def\freeue{\psi^{\left(re\right)}}%
\global\long\def\Uent{\psi_{e}^{\left(re\right)}}%
\global\long\def\Ucd{\psi_{cd}^{\left(re\right)}}%
\global\long\def\Ubond{\psi_{b}^{\left(re\right)}}%

\global\long\def\freech{\psi_{ch}^{\left(re\right)}}%

\global\long\def\boltzmann{k_{b}}%
\global\long\def\temp{T}%
\global\long\def\rat{\rho}%
\global\long\def\ratref{\rho_{0}}%
\global\long\def\lang{\beta}%
\global\long\def\langref{\beta_{0}}%
\global\long\def\ues{N_{0}}%

\global\long\def\force{f}%
\global\long\def\chforce{f_{ch}}%
\global\long\def\cdforce{f_{cd}}%
\global\long\def\bforce{f_{b}}%
\global\long\def\deltach{\delta_{ch}}%

\global\long\def\stiffness{\eta}%
\global\long\def\bstiffness{\eta_{b}}%
\global\long\def\cdstiffness{\eta_{cd}}%
\global\long\def\bstiffness{\eta_{b}}%
\global\long\def\totstiffness{\tilde{\eta}}%

\global\long\def\cdfrac{\phi_{cd}}%
\global\long\def\chfrac{\phi_{R}}%
\global\long\def\luet{L_{t}}%

\global\long\def\orienf{f}%
\global\long\def\cdvol{V_{cd}}%

\title{\vspace{-80pt}Stiffness by design in biological fibers: the influence
of microstructure and temperature}
\author{Eli Yovel $^{\footnotesize{a}}$ and Noy Cohen $^{\footnotesize{a}} \footnote{e-mail address: noyco@technion.ac.il}\,\,$
\\
$^{a}${\footnotesize{Department of Materials Science and Engineering, Technion  - Israel Institute of Technology, Haifa 3200003, Israel}}}
\maketitle
\begin{abstract}
Biologically derived fibers exhibit exceptional mechanical properties
such as high stiffness, toughness, and elasticity. The stiffness of
bio-fibers is governed by a hierarchical microstructure that is highly
sensitive to the extraction method, post-processing, and environmental
factors such as temperature. Commonly, the microstructure features
an amorphous matrix comprising polypeptide chains that interact through
weak intermolecular bonds and interconnect via crystalline domains.
In this work, we develop a microscopically motivated energy-based
model that sheds light on the underlying mechanisms governing the
stiffness of bio-fibers. Specifically, the initial deformation is
driven by (1) the entropic extension of polypeptide chains, (2) the
elastic stretching and rotation of rigid crystalline domains, and
(3) the distortion of weak intermolecular interactions, which lead
to the relative sliding of polypeptide chains. The model captures
the influence of key physical microstructural quantities such as chain
alignment, initial chain stretch, intermolecular bond strength, and
crystallite size on the overall stiffness. The merit of the model
is demonstrated through a comparison to spider silk and cocoon silk
fibers. We also employ the model to show that the reeling speed during
the extraction of spider silk fibers governs the microstructure and,
therefore, leads to different stiffnesses. We follow with a parametric
analysis that sheds light on how different microstructural quantities
affect the stiffness. Lastly, the framework is extended to account
for thermally-induced microstructural changes and the model predictions
are compared to experimental data on the stiffness of cocoon silk
fibers as a function of temperature. The findings from this work delineate
the role of microstructure on the overall stiffness and offer a pathway
for the efficient design of tunable and optimized biomimetic fibers
for target applications.
\end{abstract}

\begin{comment}
\subsubsection*{Statement of Significance}

This work addresses a critical gap in the engineering of biological
fibers by introducing a microscopically motivated, energy-based model
that sheds light on the influence of key microstructural parameters
on stiffness. By explicitly incorporating localized deformation kinematics
and thermal effects, the model is used to capture the stiffness of
diverse protein-based fibers, such as spider and cocoon silk, under
different processing conditions and temperature regimes. This framework
provides new insights into the role of key microstructural quantities
on the overall stiffness, thereby paving the way to the design of
tunable and optimized biomimetic fibers with targeted mechanical profiles
that can be used in various applications. 
\end{comment}

\section{Introduction}

Natural fibers have been used for centuries across a broad spectrum
of structural and engineering applications, most notably as reinforcing
agents in construction \citep{Ramamoorthy2019}. Over the past decades,
many of these fibers were systematically replaced by synthetic alternatives,
which are characterized by highly consistent, predictable, and exceptional
mechanical properties \citep{Ramamoorthy2019,Everitt2013}. Synthetic
fibers have a few main drawbacks. These include the high energy required
for manufacturing and processing and the sustainability aspect, with
significant technical and economic challenges that are associated
with recycling \citep{Gorreta2025,Everitt2013}. As a consequence
of the latter, synthetics are frequently disposed of in landfills
or incinerated, resulting in severe and unsustainable environmental
degradation \citep{Gorreta2025,Vinod2020}. This ecological burden
is one of the main factors that has sparked a growing global initiative
to increase the use of reliable, sustainable natural fibers with competitive
mechanical performance \citep{Li2021,Li2023,Bowen2018}. 

Protein-based bio-fibers, such as spider silk and cocoon silk, exhibit
exceptional mechanical performance, characterized by a unique combination
of high stiffness, toughness, and elasticity \citep{Xu2014,Everitt2013,Cohen2025}.
These mechanical properties are highly dependent on the microstructure,
which varies between biological species and is determined from the
extraction method, environmental factors such as temperature, and
post-processing treatments \citep{Yazawa2020,Olive2026,Olive2025,Du2006,Plaza2012}.
As a specific example, recent works show that variations in the mechanical
response of silk fibers stem from processing parameters like reeling
speed, which directly governs the initial chain pre-stretch, molecular
alignment, and the packing of the amorphous network \citep{Xu2014,Warwicker1960,Fang2016,Yazawa2025,Mazzi2014,Cohen2025}.
Therefore, it is important to fundamentally understand and systematically
quantify the relationship between the microstructure and the mechanical
performance of bio-fibers. Specifically, a robust framework that sheds
light on the influence of the network topology on mechanical properties,
and in particular stiffness, can be used to fully capitalize on the
engineering potential of these advanced bio-materials. In this context,
we point out that significant research focused on spinning and 3D
printing biomimetic, protein-based fibers is already underway \citep{Latza2015,Mu2020,Mu2021}.

Generally, polymeric bio-fibers share a common microstructure comprising
amorphous (or semi-amorphous) and crystalline domains \citep{Fratzl2007,Xu2014,Keten2010,Du2011,Shen1998,Tombolato2010}.
Within the amorphous domains, polypeptide chains interact with one
another through a series of weak intermolecular interactions, typically
hydrogen bonds, which act as local physical constraints that restrict
chain mobility \citep{Nakamura2024,Yuan2010,Roemer2008,Keten2010a,Tombolato2010}.
In addition, the chains can incorporate intramolecular interactions
that can be broken to reveal a ``hidden-length'' \citep{Du2011,Guo2017,Olive2024}.
The crystalline domains in the fiber serve as stiff, cooperative cross-links
that provide high localized stiffness and maintain the structural
integrity of the overall network \citep{Keten2010,Li2024,Tombolato2010,Shen1998}.
The specific molecular identity of the crystalline domains varies
based on the biological origin of the fiber. For instance, in spider
and cocoon silks, these domains are made of tightly packed $\beta$-sheet
nano-crystals stabilized by cooperative hydrogen bond networks \citep{Du2011,Cheng2014,Roemer2008,Xu2014,Yarger2018}.
In contrast, $\alpha$-helical protein structures govern the crystalline
reinforcement in $\alpha$-keratin fibers like human hair, wool, and
horn \citep{Tombolato2010,Yu2017}. The amorphous and the crystalline
domains and their arrangement within the fiber contribute to the overall
stiffness and toughness, and therefore govern the mechanical performance
of bio-fibers.

To determine the stiffness, one must first examine the microstructure
and the mechanisms that enable localized deformations \citep{DeTommasi2010,Olive2026,Cheng2014,Nova2010,Olive2024}.
When subjected to uniaxial extension, the global deformation of the
fiber is driven by three dominant and concurrent microstructural mechanisms:
(1) the entropic extension of the compliant polypeptide chains within
the amorphous matrix , (2) the elastic stretching and cooperative
rotation of the rigid crystalline domains, and (3) the relative sliding
of polypeptide chains, enabled by the distortion and dissociation
of weak intermolecular interactions \citep{Cohen2023,Cohen2021a}.
In highly crystalline fibers with ordered domains that dominate the
volume fraction, chain mobility is severely restricted, forcing the
system to rely on the high energetic cost of extending the cooperative
crystalline networks \citep{Keten2010,Cohen2019a,Mizushima2025,Nova2010}.
This, in turn, results in a highly stiff material. Conversely, in
fibers characterized by low initial crystallinity and a dense network
of weak physical intermolecular bonds, the initial response is compliant
due to the low energetic cost associated with bond distortion \citep{Cohen2025}.

Recent works showed that temperature also plays an important role
in the determination of the stiffness of bio-fibers \citep{Yang2005,Guan2013}.
Specifically, an increase in temperature induces a complex multi-stage
microstructural transitions within both the amorphous and crystalline
domains. At lower temperatures (typically $T\leq\text{100�}\mathrm{C}$)
the progressive expulsion of bound water molecules out of the fiber
promotes network compaction and motivates the formation of additional
intermolecular hydrogen bonds, thereby increasing the stiffness \citep{Greco2025,Mazzi2014}.
As the temperature rises, the additional thermal energy introduces
molecular vibrations that systematically weaken and dissociate intermolecular
and intramolecular interactions, thereby enhancing chain mobility
\citep{Yuan2010,Dao2017,Guan2013}. This results in a decrease in
the overall stiffness as entropic mechanisms take over enthalpic ones.
At extreme temperatures approaching thermal degradation, cooperative
breakdown extends to crystalline domains, ultimately resulting in
the loss of structural integrity of the fiber. Understanding the influence
of temperature on the microstructure and the stiffness of the fiber
is essential for the design of reliable and optimized bio-fibers operating
at different environments. 

The aim of this work is to provide a fundamental understanding into
the underlying mechanisms that govern the stiffness of bio-fibers.
To this end, we derive an energetically motivated, microstructural
model that bridges the local response, which includes localized deformation
mechanisms, to the macroscopic mechanical response and yields an expression
for the Young's modulus. We validate the model by comparing the predicted
Young's modulus to established experimental data in high-performance
bio-fibers. Next, we conduct a parametric analysis to systematically
explore how distinct microstructural quantities control the fiber
stiffness. Lastly, we extend the model to capture the effects of temperature
on the microstructural evolution. To highlight the merit of the proposed
framework, we compare the model predictions to experiments that measured
the dependence of stiffness on temperature in cocoon silk fibers. 

\section{Deformation mechanisms in bio-fibers}

Many biological fibers share a similar structure, comprising a semi-amorphous
matrix of polypeptide chains that interact with each other through
weak intermolecular bonds and interconnect through crystalline domains,
which serve as effective cross-links, to form a 3-dimensional network.
Generally, the initial deformation and the stiffness of these fibers
is governed by three dominant mechanisms: (1) the entropic extension
of the chains, (2) the deformation of the crystalline domains, and
(3) the distortion of weak intermolecular bonds. 

In the following, we describe the kinematics of the three deformation
mechanisms and propose expressions for their energies. This allows
us to develop the strain energy-density of the fiber, and consequently
compute the stress, which enables us to obtain a closed-form expression
for the stiffness (the Young's modulus) as a function of the microstructure.

\subsection{Kinematics}

\begin{figure}[t]
\begin{centering}
\includegraphics[width=12cm]{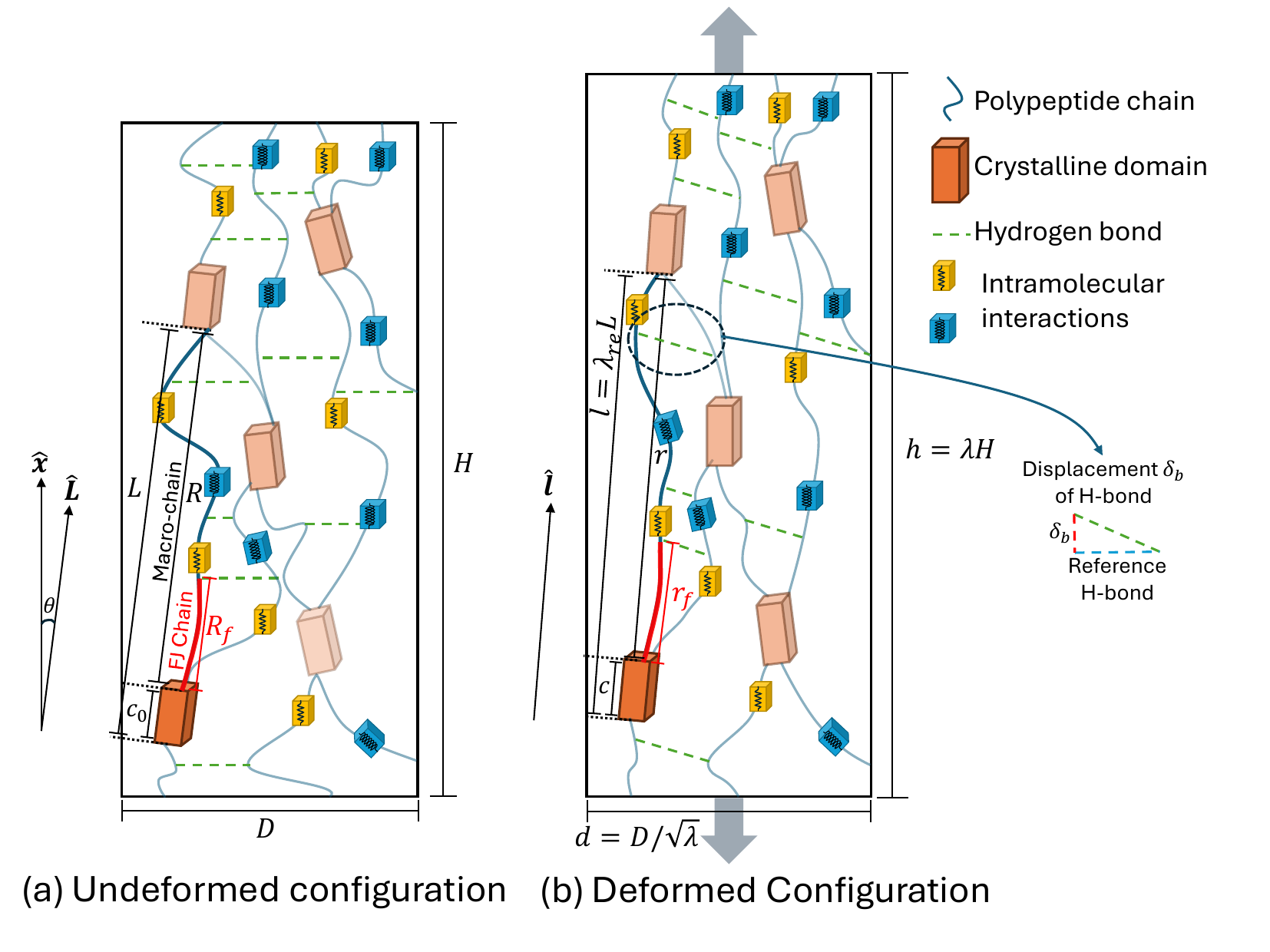}
\par\end{centering}
\caption{The microstructure of a typical bio-fiber subjected to uniaxial extension
in the (a) undeformed and (b) deformed configuration. \label{fig: Kinematic_changes}}
\end{figure}

Consider an anisotropic bio-polymer fiber with a length $H$ and a
diameter $D$ that is aligned along the $\xh$-direction in a coordinate
system $\left\{ \xh,\yh,\zh\right\} $. To determine the stiffness
of the fiber, we subject it to uniaxial extension such that its deformed
length and diameter are $h$ and $d$, respectively. The stretch,
or the ratio between the deformed and the reference lengths of the
fiber, is $\stretch=h/H$. Biological fibers are generally nearly
incompressible and, accordingly, the stretch along the transverse
directions ($\yh$ and $\zh$) is $1/\sqrt{\lambda}=d/D$ \citep{Mori2009,Guinea2006,Grubb1997}.
Therefore, the macroscopic deformation gradient can be written as 

\begin{equation}
\defgradT=\stretch\xh\otimes\xh+\frac{1}{\sqrt{\stretch}}\left(\yh\otimes\yh+\zh\otimes\zh\right).
\end{equation}

To characterize the local behavior, we consider a representative element
that is made of a crystalline domain with an initial length $\cref$
that is connected to a polypeptide chain, as shown in Fig. \ref{fig: Kinematic_changes}a.
We refer to a polypeptide chain connecting two crystalline domains
as a \textit{macro-chain}, which comprises $\links$ repeat units
of length $\linklen$ and has an end-to-end distance $\macchainref$.
The distance $\macchainref$ is also referred to as the\emph{ intercrystallite
distance}. The macro-chain interacts with its neighboring chains through
$\bonds$ weak intermolecular bonds. The bonds act as ``weak cross-links''
that divide the macro-chain into $\bonds+1$ \textit{freely jointed
}\emph{chains} with $\links/\left(\bonds+1\right)$ repeat units.
 On average, the initial length of a freely jointed chain is $\chainref=\macchainref/\left(\bonds+1\right)$.
We conjecture that the crystalline domain, the macro-chain, and the
freely jointed chains are aligned along the same direction $\dirTref$
(see Fig. \ref{fig: Kinematic_changes}a), and accordingly the length
of a representative element is
\begin{equation}
\lueref=\left(\bonds+1\right)\chainref+\cref.\label{eq:ref_len}
\end{equation}

Next, we aim to understand the response of a representative element
in a fiber that is subjected to uniaxial extension. To this end, consider
a representative element that forms an angle $0\le\ang\le\pi$ with
the fiber direction such that $\dirTref\cdot\xh=\cos\ang$ (see Fig.
\ref{fig: Kinematic_changes}a). In the deformed state, the length
and the direction of this element are $\luecur$ and $\dirTcur$,
respectively, as shown in Fig. \ref{fig: Kinematic_changes}b. Following
common practice from polymer physics \citep{Flory1953,Treloar1975},
we assume that the representative elements experience affine deformations
(i.e. $\luecur\,\dirTcur=\defgradT\left(\lueref\dirTref\right)$)
such that 
\begin{equation}
\luecur=\stretchue\lueref,
\end{equation}
where 
\begin{equation}
\stretchue=\sqrt{\defgradT\dirTref\cdot\defgradT\dirTref}=\sqrt{\frac{\stretch^{3}+1+\left(\stretch^{3}-1\right)\cos\left(2\ang\right)}{2\stretch}},\label{eq:stretch_ue}
\end{equation}
is the stretch of the representative element. The deformed direction
of the representative element is $\dirTcur=\defgradT\dirTref/\stretchue$.

Eq. \ref{eq:stretch_ue} reveals that under uniaxial extension the
stretch of a representative element $\stretchue$ depends on its orientation
with respect to the fiber direction $\xh$. Specifically, elements
that are aligned along the fiber direction (i.e. $\ang=0$) extend
while elements that are perpendicular to $\xh$ $\left(\ang=\pi/2\right)$
shorten. 

Fig. \ref{fig: Kinematic_changes}b illustrates the deformed configuration
of the element. To characterize the local deformations, we denote
the deformed length of the crystalline domain and the end-to-end distance
of the polypeptide macro-chain by $\ccur$ and $\macchaincur$, respectively.
The freely jointed chains along the macro-chain are characterized
by an average end-to-end distance $\chaincur=\macchaincur/\left(\bonds+1\right)$.
The weak intermolecular interactions distort such that their contribution
to the overall increase in length of the representative element is
$\bcur$, as highlighted in Fig. \ref{fig: Kinematic_changes}b. Accordingly,
the deformed length of the representative element is 
\begin{equation}
\luecur=\stretchue\lueref=\left(\bonds+1\right)\chaincur+\ccur+\bcur.\label{eq:cur_len}
\end{equation}

Before proceeding, it is worth mentioning that the size of the crystalline
domain relative to the macro-chain is key in determining the fiber
stiffness. Specifically, in the case of fibers with a high degree
of crystallinity ($\cref\gg\macchainref$), the entropic extensibility
and mobility of the chains is restricted by the crystalline domains.
Recall that these domains have an ordered and organized structure,
commonly comprising stacked $\beta$-sheets (as in cocoon and spider
silks \citep{Xu2014,Keten2010}) or $\alpha$-helices (as in human
hair and horn \citep{Yu2017,Harland2022,Kitchener1987,Tombolato2010}),
which provide stiffness through cooperative bonding \citep{Rossmeisl2004,Koch2005}.
Therefore, the deformation of the fiber is enabled by the cooperative
rotation and slight extension of the crystalline domains, resulting
in a stiff fiber with a high Young's modulus and limited extensibility
\citep{Keten2010,Mizushima2025,Nova2010,DeTommasi2010}. In typical
bio-fibers with low crystallinity and a high degree of intermolecular
interactions ($\bonds\gg0$), the fiber is initially compliant due
to the low energetic cost of distorting weak intermolecular bonds.

\subsection{The energy of a representative element \label{subsec:energy_RE}}

To determine the stiffness of a bio-fiber, we first model the constitutive
response of a representative element. The free energy of a representative
element $\freeue\left(\stretch,\dirTref\right)$ depends on its orientation
with respect to the fiber axis and comprises three contributions -
(1) the energy stemming from the entropic deformation of freely jointed
chains in the polypeptide chain $\Uent\left(\links,\bonds,\chaincur\left(\stretch\right)\right)$,
(2) the deformation energy of the crystalline domains $\Ucd\left(\ccur\right)$,
and (3) the energy due to the distortion of the intermolecular bonds
$\Ubond\left(\bcur\right)$. Accordingly, the total energy of a representative
element is
\begin{equation}
\freeue\left(\stretch,\dirTref\right)=\Uent+\Ucd+\Ubond.
\end{equation}

The free energy of a freely jointed chain with $\links/\left(\bonds+1\right)$
repeat units can be written as \citep{Flory1953,Treloar1975}

\begin{equation}
\freech\left(\links,\bonds,\chaincur\left(\stretch\right)\right)=\boltzmann T\frac{\links}{\bonds+1}\left(\rat\,\lang\left(\rat\right)+\ln\frac{\lang\left(\rat\right)}{\sinh\left(\lang\left(\rat\right)\right)}\right),\label{eq:free_energy_FJC}
\end{equation}
where $\boltzmann$ is the Boltzmann constant, $T$ is the temperature,
$\rat=\chaincur\left(\bonds+1\right)/\links\,\linklen<1$ is the ratio
between the end-to-end distance and the contour length of the chain,
and $\lang\left(\rat\right)$ is determined from the Langevin function
$\rat=\coth\lang-1/\lang$. Note that the since $\macchaincur=\chaincur\left(\bonds+1\right)$,
the ratio $\rat$ of the freely jointed chain and the macro-chain
are identical. Since there are $\bonds+1$ freely jointed chains in
a macro-chain, we can write 
\begin{equation}
\Uent\left(\links,\bonds,\chaincur\left(\stretch\right)\right)=\left(\bonds+1\right)\freech,\label{eq:FE_macro_chain}
\end{equation}
where $\freech$ is given in Eq. \ref{eq:free_energy_FJC}. 

Generally, the force required to maintain an end-to-end distance $\chaincur$
is $\chforce=\partial\freech/\partial\chaincur=\boltzmann T\lang/\linklen$
\citep{Kuhn1942,Treloar1975}. Recall that the referential freely
jointed chain is characterized by an end-to-end distance $\chaincur=\chainref>0$
and therefore experiences an initial non-zero force which is typically
balanced by the other chains in the network. The increase in force
required to stretch the freely jointed chain from its initial ratio
$\ratref=\chainref\left(\bonds+1\right)/\links\,\linklen$ to a ratio
$\rat=\chaincur\left(\bonds+1\right)/\links\,\linklen$ can be approximated
through a Taylor series around $\rat=\ratref$,

\begin{equation}
\chforce=\frac{\boltzmann\temp}{\linklen}\left(\lang\left(\rat\right)-\langref\right)\approx\frac{\boltzmann\temp}{\linklen}\langref'\cdot\left(\rat-\ratref\right),\label{eq:force_chain}
\end{equation}
where $\langref=\lang\left(\ratref\right)$ and 
\begin{equation}
\langref'=\frac{\d\lang}{\d\rat}\left(\rat=\ratref\right)=\frac{\langref^{2}\sinh^{2}\langref}{\sinh^{2}\langref-\langref^{2}}.\label{eq:lang_derivative}
\end{equation}

Next, we model the crystalline domains as linear springs with a spring
constant $\cdstiffness\boltzmann T/\linklen^{2}$, where $\cdstiffness$
is the dimensionless stiffness to stretching. The deformation energy
of a crystalline domain is 
\begin{equation}
\Ucd\left(\ccur\left(\stretch\right)\right)=\frac{\cdstiffness\boltzmann T}{2\linklen^{2}}\left(\ccur-\cref\right)^{2},
\end{equation}
and the stretching force is
\begin{equation}
\cdforce=\frac{\cdstiffness\boltzmann T}{\linklen^{2}}\left(\ccur-\cref\right).\label{eq:force_CD}
\end{equation}

Lastly, the intermolecular bonds are modeled as distortional linear
springs with a spring constant $\bstiffness\boltzmann T/\linklen^{2}$,
where $\bstiffness$ is the dimensionless distortional stiffness.
Since there are $\bonds$ bonds that characterize the interaction
between two chains, the free energy of the intermolecular bonds in
a representative element can be written as
\begin{equation}
\Ubond\left(\bcur\left(\stretch\right)\right)=\frac{\bonds}{2}\frac{\bstiffness\boltzmann T}{2\linklen^{2}}\bcur^{2}.\label{eq:bond_energy}
\end{equation}
The distortional force on a single bond is
\begin{equation}
\bforce=\frac{\bstiffness\boltzmann T}{\linklen^{2}}\bcur.\label{eq:force_bond}
\end{equation}

Before proceeding, we remark that the order of magnitude of the $\bstiffness$
can be estimated for different intermolecular interactions. For example,
the dissociation energy of hydrogen bonds in peptides is $\sim5-10\boltzmann T$
\citep{Sheu2003}. The typical length of a repeat unit is $\linklen\sim\mathrm{nm}$
and if we consider the displacement $\bcur\approx\linklen$ at failure,
one obtains $\bstiffness\sim10-20$. It is worth mentioning that typically
the crystalline domains are significantly stiffer than the intermolecular
bonds \citep{Keten2010,Nova2010,DeTommasi2010}, and therefore $\cdstiffness\gg\bstiffness$.

\begin{figure}
\hspace*{\fill}\includegraphics[width=12cm]{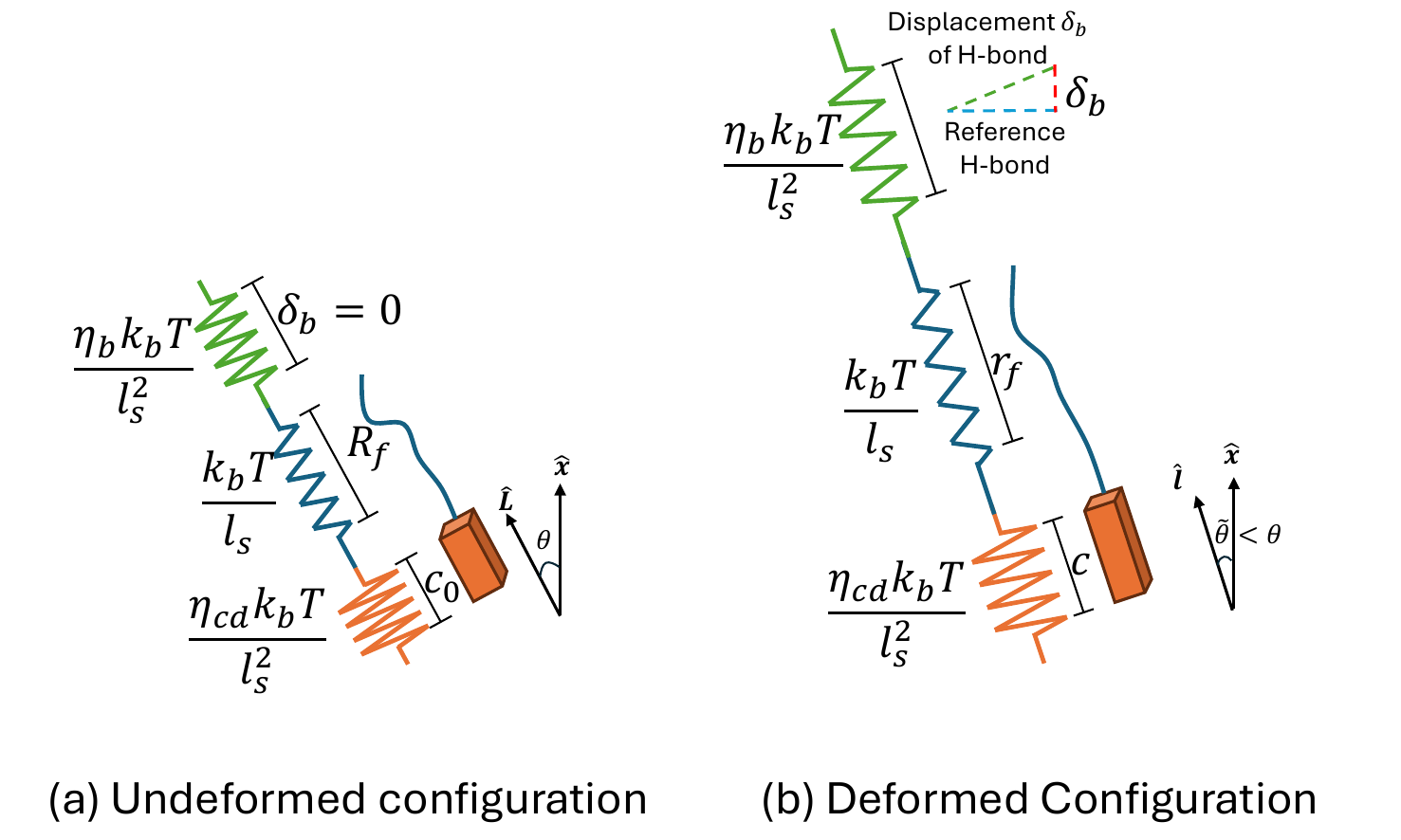}\hspace*{\fill}

\caption{The deformation mechanisms in a bio-fiber as a series of springs in
(a) undeformed and (b) deformed configurations. \label{fig:springs_changes}}
\end{figure}

To determine the extension and the distortion of the three constituents
in a representative element, we treat these as a series of springs
that are connected in series and subjected to a force $f$. The undeformed
and the deformed system of springs are depicted in Figs. \ref{fig:springs_changes}a
and \ref{fig:springs_changes}b, respectively. The macro-chain, the
crystalline domain, and the bonds experience the same force
\begin{equation}
f=\chforce=\cdforce=\bforce.\label{eq:force_equilibrium}
\end{equation}

The forces in Eqs. \ref{eq:force_chain}, \ref{eq:force_CD}, and
\ref{eq:force_bond}, the equilibrium condition in Eq. \ref{eq:force_equilibrium},
and the kinematic relations in Eqs. \ref{eq:stretch_ue} and \ref{eq:cur_len},
along with the definitions $\rat=\macchaincur/\links\linklen$ and
$\ratref=\macchainref/\links\,\linklen$ comprise a set of equations
that can be used to determine the deformed intercrystallite distance
$\macchaincur$, the deformed length of the crystalline domain $\ccur$,
and the displacement of the intermolecular bonds $\bcur$, 
\begin{align}
\macchaincur= & \macchainref+\frac{\cdstiffness\bstiffness}{\totstiffness}\links\lueref\left(\stretchue-1\right),\\
\ccur= & \cref+\frac{\bstiffness\langref'}{\totstiffness}\lueref\left(\stretchue-1\right),\\
\bcur= & \frac{\cdstiffness\langref'}{\totstiffness}\lueref\left(\stretchue-1\right),
\end{align}
where $\totstiffness=\links\,\cdstiffness\bstiffness+\langref'\left(\cdstiffness+\bstiffness\right)$
and $\langref'$ is given in Eq. \ref{eq:lang_derivative}.

\subsection{Distribution of representative elements in the network}

\global\long\def\odf{p}%
\global\long\def\odfpar{b}%

To account for the orientational distribution of the representative
elements in the network, an orientational distribution function $\odf\left(\theta\right)$
that accounts for the probability of having a representative element
with the orientation $\ang$ must be defined. In this work, we employ
the orientation distribution function
\begin{equation}
\odf\left(\theta\right)=4\sqrt{\frac{\odfpar}{2\pi}}\frac{\exp\left(\odfpar\left(\cos\left(2\ang\right)+1\right)\right)}{\mathrm{erfi}\left(\sqrt{2\odfpar}\right)},\label{eq:odf}
\end{equation}
where $0\le\ang\le\pi$ is the angle between the direction $\dirTref$
and the fiber direction $\xh$, $\mathrm{erfi}\left(x\right)$ is
the well-known imaginary error function, and $\odfpar$ is a parameter
that accounts for the degree of anisotropy. The orientational distribution
function $\odf\left(\ang\right)$ is periodic in $\pi$ and satisfies
$\int_{0}^{\pi}\odf\sin\ang/2\,\d\ang=1$, where $\sin\ang/2\,\d\ang$
is the normalized solid angle for a unit sphere. 

In Eq. \ref{eq:odf}, the limit $\odfpar\rightarrow0$ corresponds
to a network with randomly oriented and uniformly distributed representative
elements. Alternatively, $\odfpar\rightarrow\infty$ represents a
highly aligned network in which all representative elements are aligned
along the fiber direction $\xh$ (i.e. characterized by the angles
$\ang=0$ and $\ang=\pi$). 

\subsection{Stiffness of a bio-fiber \label{subsec:Stiffness-equation}}

\global\long\def\freeg{\tilde{\psi}}%
\global\long\def\free{\psi}%
\global\long\def\piola{P_{x}}%
\global\long\def\stress{\sigma_{x}}%

\global\long\def\shear{\mu}%
\global\long\def\normL{\xi}%
\global\long\def\cdfrac{\phi_{cd}}%
\global\long\def\cdvol{V_{cd}}%

The total energy-density of a bio-fiber with a representative element
density $\ues$ is 
\begin{equation}
\freeg\left(\stretch\right)=\ues\sum\freeue\left(\stretch,\ang\right)\odf\left(\ang\right)=\ues\int_{0}^{\pi}\freeue\left(\stretch,\ang\right)\odf\left(\ang\right)\frac{\sin\ang}{2}\d\ang,
\end{equation}
where the summation on the left hand-side is carried over all representative
elements and is replaced in passing by an integral. Before the stiffness
can be determined, recall that a bio-fiber network with a pre-aligned
distribution of chains is characterized by $\odfpar>0$. In such cases,
an inherent internal stress that maintains the position of the polypeptide
chains develops. To account for the residual stress in the reference
traction free configuration, we ``shift'' the energy-density function
such that 
\begin{equation}
\free\left(\stretch\right)=\freeg\left(\stretch\right)-\freeg'\left(1\right)\left(\stretch-1\right),
\end{equation}
where $\freeg'\left(1\right)=\d\freeg/\d\stretch\left(\stretch=1\right)$
is the residual nominal stress in the reference configuration. Recall
that in isotropic materials there is no residual stress and therefore
$\freeg'\left(1\right)=0$. 

Subsequently, the stress can be written as $\stress=\freeg'\left(\stretch\right)-\freeg'\left(1\right)$
and the Young's modulus is computed via
\begin{equation}
E=\free''\left(1\right)=\shear\normL\frac{\bstiffness\cdstiffness}{8\totstiffness}\left(3\links\langref\funcone\left(\odfpar\right)+\frac{\normL\langref'}{4\totstiffness}\left(\langref'\left(2\bstiffness+\bonds\cdstiffness\right)+2\links\bstiffness\cdstiffness\right)\functwo\left(\odfpar\right)\right),\label{eq:Young_modulus}
\end{equation}
where $\langref'$ is given in Eq. \ref{eq:lang_derivative}, $\shear=\ues\boltzmann T$,
and $\normL=\lueref/\linklen$. The functions 
\begin{equation}
\funcone\left(\odfpar\right)=2-\frac{9}{8\odfpar^{2}}-\frac{1}{\odfpar}+\frac{\left(9-4\odfpar\right)\exp\left(2\odfpar\right)}{\sqrt{8\pi}\odfpar^{3/2}\mathrm{erfi}\left(\sqrt{2\odfpar}\right)},
\end{equation}
and 
\begin{equation}
\functwo\left(\odfpar\right)=4+\frac{27}{4\odfpar^{2}}+\frac{6}{\odfpar}-3\frac{\left(9-4\odfpar\right)\exp\left(2\odfpar\right)}{\sqrt{2\pi}\odfpar^{3/2}\mathrm{erfi}\left(\sqrt{2\odfpar}\right)},
\end{equation}
capture the dependence of the modulus on the orientational distribution
of the representative elements in the fiber. 

The model parameters in Eq. \ref{eq:Young_modulus} for the stiffness
are physically motivated and can be measured or estimated. Specifically,
experiments and MD simulations can provide estimates for the stiffness
of the crystalline domains $\cdstiffness$ and the intermolecular
bonds $\bstiffness$. Available measurements of the intercrystallite
distance $\macchainref$ and the crystalline domain size $\cref$
are used to determine the parameter $\normL$ and estimate $\ratref$.
The parameter $\shear$ and the characteristics of the amorphous domain,
such as the number of repeat units  $\links$ and the number of intermolecular
bonds $\bonds$, are fitted to the experimental data. The orientational
distribution of the representative elements, characterized by $\odfpar$,
can be determined from the azimuthal FWHM (Full Width at Half Maximum)
of diffraction peaks, where a narrower peak width indicates a higher
degree of molecular orientation along the fiber axis, as shown by
\citet{Du2006}. For convenience, we list the model parameters and
their significance in Table \ref{tab:parameters_table}.

\begin{table}
\caption{Model parameters and their physical significance \label{tab:parameters_table}}

\hspace*{\fill}%
\begin{tabular}{cc}
\toprule 
Parameter & Physical significance\tabularnewline
\midrule
\midrule 
$\bonds$ & Number of intermolecular physical bonds\tabularnewline
\midrule 
$\links/\linklen$ & Number / length of repeat units\tabularnewline
\midrule 
$\macchainref/\macchaincur$ & Undeformed / deformed end-to-end distance of macro-chain\tabularnewline
\midrule 
$\chainref/\chaincur$ & Undeformed/deformed end-to-end distance of freely jointed chain\tabularnewline
\midrule 
$\lueref/\luecur$ & Undeformed/deformed length of representative element\tabularnewline
\midrule 
$\cref/\ccur$ & Undeformed/deformed length of crystalline domain\tabularnewline
\midrule 
$\normL$ & Normalized length of a representative element\tabularnewline
\midrule 
$\ratref/\rat$ & Undeformed/deformed ratio between the end-to-end distance and the
contour length\tabularnewline
\midrule 
$\odfpar$ & \begin{cellvarwidth}[t]
\centering
Alignment of representative elements:

$\odfpar\rightarrow0$ - isotropic distribution

$\odfpar\rightarrow\infty$ - highly aligned network
\end{cellvarwidth}\tabularnewline
\midrule 
$\bcur$ & Displacement of intermolecular bonds\tabularnewline
\midrule 
$\ues$ & Density of representative element\tabularnewline
\midrule 
$\stiffness_{i}$ & Stiffness of element $i=b,cd$\tabularnewline
\bottomrule
\end{tabular}\hspace*{\fill}
\end{table}

\section{Comparison to experiments}

To illustrate the merit of the proposed framework, we compare its
predictions to experimental findings. First, we focus on spider silk
and cocoon silk, which are two semi-crystalline bio-polymeric fibers
with a similar microstructure, comprising crystalline domains made
of stacked $\beta$-sheets and amorphous domains with polypeptide
chains that are connected through weak intermolecular bonds. Experiments
have shown that these fibers are anisotropic due to an inherent chain
alignment induced during the spinning process \citep{Eles2004,Yazawa2020,Qi2017,Du2011,Cheng2014,Roemer2008,Xu2014,Yarger2018}.
Secondly, we examine the work of \citet{Du2006}, which measured the
stiffness of \emph{Nephila pilipes} spider silk fibers retrieved at
different reeling speeds, and employ the proposed framework to understand
the effects of reeling speeds on the microstructure. 

\subsection{Young modulus of bio-fibers \label{subsec:Comparison_bio_fibers}}

We begin by examining spider silk and cocoon silk and estimating the
stiffness based on Eq. \ref{eq:Young_modulus}. These structures comprise
intermolecular hydrogen bonds that connect polypeptide chains \citep{Qi2017,Liu2022}.
Following the discussions in Sec. \ref{subsec:energy_RE}, the stiffness
of the intermolecular hydrogen bonds and the crystalline domains are
set to $\bstiffness=10$ and $\cdstiffness=100$, respectively. In
addition, we set the characteristic Kuhn length $\linklen=1\,\mathrm{nm}$.
The remaining model parameters are estimated based on experimental
characterizations, as described in the following. 

We begin with the spider silk fiber. The crystallinity of spider silk
fibers ranges between $22-27\%$ \citep{Yazawa2020,Xu2014,Du2011}
and the dimensions of a single crystal are $2-5\times2-2.5\times6-7\,\mathrm{nm}$,
where $a\times b\times c$ are the lengths along the side chain, the
hydrogen bond, and the crystallite directions, respectively \citep{Roemer2008,Xu2014,Yarger2018}.
The ratio $\normL=\left(\macchainref+\cref\right)/\linklen=28.8$
can be computed from the intercrystallite distance $\macchainref=22.3\,\mathrm{nm}$
\citep{Du2011,Du2006} and the length of the crystallite along the
direction of the macro-chain $\cref=6.5\,\mathrm{nm}$ \citep{Roemer2008,Xu2014,Yarger2018}.
The amorphous domains are characterized by $\links=33$ and $\bonds=25$,
leading to an initial chain extension characterized by $\ratref=0.68$,
which accounts for the initial extensibility stemming from the spinning
process. We also set the parameter $\shear=20\,\mathrm{MPa}$. 

To estimate the orientation parameter $\odfpar$, we follow previous
works that employed X-ray diffraction along with Hermans orientation
function
\begin{equation}
\orienf=\frac{1}{2}\left(3\langle\cos^{2}\ang\rangle-1\right),\label{eq:orientation f}
\end{equation}
where
\begin{equation}
\langle\cos^{2}\ang\rangle=\int_{0}^{\pi}\cos^{2}\ang\,\odf\left(\ang\right)\frac{\sin\ang}{2}\d\ang,\label{eq:cos_sq}
\end{equation}
denotes the average of $\cos^{2}\ang$. Recall that $\ang$ is the
angle of a representative element with respect to the fiber axis.
Once $\orienf$ is known, one can calculate $\langle\cos^{2}\ang\rangle$
and solve Eq. \ref{eq:cos_sq} for the alignment parameter $\odfpar$,
which appears in the orientation distribution function $\odf\left(\ang\right)$
(Eq. \ref{eq:odf}). At a reeling speed of $10\,\mathrm{mm/s}$, \citet{Du2006}
and \citet{Xu2014} reported $\orienf$ in the range of $0.94-0.97$,
which corresponds to $\odfpar\sim13-25$. Accordingly, we set $\odfpar=19$. 

A similar process is carried out for cocoon silk. The literature reports
a crystallinity of $45-55\%$ \citep{Kim2023,Du2011,Xu2014,Keten2010}
with crystalline size $2.1-2.6\times2.7-3.5\times6.4-11.6\,\mathrm{nm}$
\citep{Xu2014,Du2011,Cheng2014}. The intercrystallite distance $\macchainref=8.7\,\mathrm{nm}$
\citep{Du2011} enables us to estimate $\normL=17.7$ with $\cref=9\,\mathrm{nm}$.
Next, we fit $\shear=30\,\mathrm{MPa}$ and the number of repeat units
and intermolecular bonds $\links=12$ and $\bonds=6$, respectively,
corresponding to a ratio $\ratref=0.725$. Note that with this definition
the chains in the cocoon silk are slightly more stretched than those
in spider silk. The alignment parameter $\odfpar=13.7$ is determined
from Eqs. \ref{eq:orientation f} and \ref{eq:cos_sq} with $\orienf=0.944$
at reeling speeds of $13\,\mathrm{mm/s}$ \citep{Xu2014}. 

\begin{table}
\caption{The model parameters characterizing the bio-fibers spider silk and
cocoon silk. Parameters marked with a superscript {*} are computed
via experimentally measured quantities. \label{tab:E_comparison_properties}}

\hspace*{\fill}%
\begin{tabular}{>{\centering}p{5cm}>{\centering}p{3cm}>{\centering}p{3cm}}
\toprule 
 & Spider silk & Cocoon silk\tabularnewline
\midrule
\midrule 
$\shear$ & $20\,\mathrm{MPa}$ & $30\,\mathrm{MPa}$\tabularnewline
\midrule 
$\normL${*} & $28.8$ & $17.7$\tabularnewline
\midrule 
Alignment parameter $\odfpar${*} & $19$ & $13.7$\tabularnewline
\midrule 
Kuhn  $\links$ & $33$ & $12$\tabularnewline
\midrule 
Intermolecular hydrogen bonds $\bonds$ & $25$ & $6$\tabularnewline
\midrule 
Ratio $\ratref${*} & $0.68$ & $0.725$\tabularnewline
\midrule 
$E$ (Eq. \ref{eq:Young_modulus}) & $6.4\,\mathrm{GPa}$ & $10.4\,\mathrm{GPa}$\tabularnewline
\midrule 
Experimentally measured $E$  & $3-10\,\mathrm{GPa}$ & $4.3-17.4\,\mathrm{GPa}$\tabularnewline
\bottomrule
\end{tabular}\hspace*{\fill}
\end{table}

The Young's moduli of the two fibers are computed with Eq. \ref{eq:Young_modulus}.
The model parameters and the resulting stiffness are summarized in
Table \ref{tab:E_comparison_properties}, where the parameters directly
obtained from experimental findings are marked with a superscript
{*}. It is important to note that within the literature there are
large deviations in the reported stiffness values of the fibers. These
variations are attributed to the variability of biological materials,
as well as other conditions such as the spinning conditions, the age,
the gender, and the diet of the animal. For spider silk fibers we
find $E=6.4\,\mathrm{GPa}$, which falls within the range of values
reported in the literature $3-10\,\mathrm{GPa}$ \citep{Yarger2018,Plaza2012,Yazawa2020,Brookes2008}.
In cocoon silk fibers the calculated stiffness is $E=10\,.4\mathrm{GPa}$,
which corresponds to the experimentally measured range of stiffness
values $4.3-17.4\,\mathrm{GPa}$ \citep{Yarger2018,Yazawa2025,Colomban2012,Fu2009,Guan2025,Chen2019}
.

The above framework reveals that the stiffness of the fibers is governed
by the crystallinity, which is accounted for through the parameter
$\normL$ and the ratio $\ratref$, the density of representative
elements, characterized by $\shear$, the initial extension of the
chains $\ratref$, the intermolecular hydrogen bonds $\bonds$, and
the network alignment, characterized by $\odfpar$. It should be noted
that the parameter $\shear$, which characterizes the shear modulus
of rubbery networks, is in the order of $10\,\mathrm{MPa}$ while
the stiffness of the fibers is in $\mathrm{GPa}$. Therefore, the
stiffness of the fibers mostly stems from the crystallinity, the alignment
of the polypeptide chains, and the high density of intermolecular
bonds. These mechanisms are similar to those observed in glassy polymers. 

\subsection{The effect of microstructure on $E$}

\global\long\def\reelingsp{s}%

To validate the ability of the model to capture the response of bio-fibers
with different microstructures, we compare its predictions to the
experiments of \citet{Du2006} on \emph{Nephila}\emph{ pilipes} spiders
retrieved at different reeling speeds $\reelingsp=1,2.5,10,25$, and
$100\,\mathrm{mm/s}$. This work shows that the reeling speed greatly
influences the microstructure of the fiber and, accordingly, its mechanical
properties. As in the previous section, we set $\bstiffness=10$,
$\cdstiffness=100$, and $\linklen=1\,\mathrm{nm}$ and estimate the
remaining model parameters through experimental data. 

\begin{figure}
\begin{centering}
(a)\includegraphics[width=4.5cm]{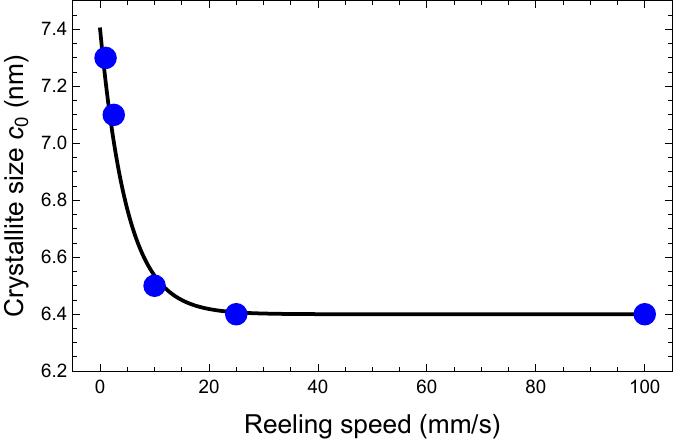}~~(b)\includegraphics[width=4.5cm]{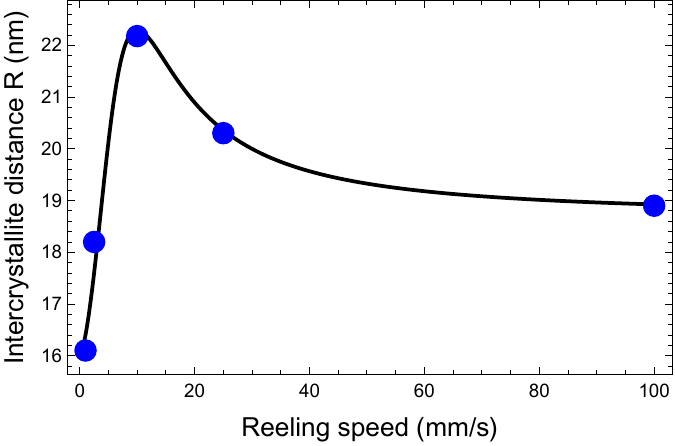}~~(c)\includegraphics[width=4.5cm]{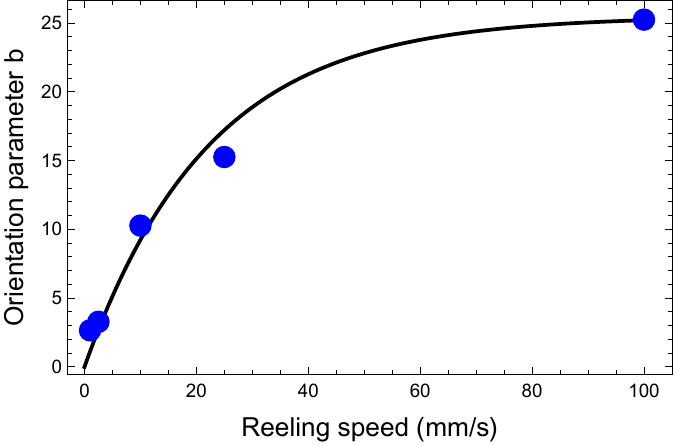}
\par\end{centering}
\caption{(a) The crystallite size $\protect\cref$ (Eq. \ref{eq:c0_eq}), (b)
the end-to-end distance of a macro-chain $\protect\macchainref$ (Eq.
\ref{eq:R_eq}), and (c) the orientation parameter $\protect\odfpar$
(Eq. \ref{eq:b_eq}) as a function of the reeling speed $\protect\reelingsp$.
The circle marks denote the experimental results of \citet{Du2006}.
\label{fig:du_exp_model_parameters}}
\end{figure}

We begin with the estimation of the normalized length $\normL$. \citet{Du2006}
reported the length $\cref$ along the direction of the crystallite.
This value exponentially decreases from $7.3\,\mathrm{nm}$ at reeling
speed of $1\,\mathrm{mm/s}$ to $6.4\,\mathrm{nm}$ at $100\,\mathrm{mm/s}$.
Accordingly, we propose the form
\begin{equation}
\cref\left(\reelingsp\right)=\left(\tilde{\cref}-\bar{\cref}\right)\exp\left(-\tau_{\cref}\reelingsp\right)+\bar{\cref},\label{eq:c0_eq}
\end{equation}
where $\tilde{\cref}=7.4\,\mathrm{nm}$ is the size at reeling speed
$\reelingsp=0$, $\bar{\cref}=6.4\,\mathrm{nm}$ is the length at
$\reelingsp\rightarrow\infty$, and $\tau_{\cref}=0.2$ is a rate
constant. The agreement between Eq. \ref{eq:c0_eq} and the experimental
results is shown in Fig. \ref{fig:du_exp_model_parameters}a.

The intercrystallite length along the meridional $\macchainref_{m}$
and the equatorial $\macchainref_{e}$ directions was also measured
and reported by \citet{Du2006}. In this work, we estimate the end-to-end
distance of a macro-chain via $\macchainref=\sqrt{\macchainref_{m}^{2}+\macchainref_{e}^{2}}$.
To capture the dependence of $\macchainref$ on the reeling speed
$\reelingsp$, we propose the expression 
\begin{equation}
\macchainref\left(\reelingsp\right)=\macchainref_{\infty}+\frac{1+2\eta_{1}\reelingsp_{t}^{\alpha}}{1+\eta_{2}\reelingsp_{t}^{2\alpha}}\left(\macchainref_{0}-\macchainref_{\infty}\right).\label{eq:R_eq}
\end{equation}
Here, $\macchainref_{0}=\macchainref\left(\reelingsp\rightarrow0\right)=16\,\mathrm{nm}$
and $\macchainref_{\infty}=\macchainref\left(\reelingsp\rightarrow\infty\right)=18.7\,\mathrm{nm}$
denote the intercrystallite distances in the limit of slow and fast
reeling speeds, respectively, and we define the maximum value $\macchainref_{p}=\macchainref\left(\reelingsp=\reelingsp_{p}\right)=22.3\,\mathrm{nm}$
at $\reelingsp=\reelingsp_{p}=10\,\mathrm{mm/s}$. Furthermore, $\eta_{1}=\left(\macchainref_{p}-\macchainref_{0}\right)/\left(\macchainref_{0}-\macchainref_{\infty}\right)$,
$\eta_{2}=\left(\macchainref_{p}-\macchainref_{0}\right)/\left(\macchainref_{p}-\macchainref_{\infty}\right)$,
$\reelingsp_{t}=\reelingsp/\reelingsp_{p}$, and $\alpha=1.5$ is
a constant determining the rate of decay. Fig. \ref{fig:du_exp_model_parameters}b
shows the merit of Eq. \ref{eq:R_eq} through a comparison to experimental
data. Eqs. \ref{eq:c0_eq} and \ref{eq:R_eq} are used to compute
the normalized length $\normL=\left(\macchainref+\cref\right)/\linklen$. 

Interestingly, the intercrystallite distance initially increases with
reeling speed, peaks at $\sim10\,\mathrm{mm/s}$, and then decreases.
This counterintuitive behavior is explained as follows: under low
reeling speeds, the amorphous chains are somewhat relaxed and loosely
packed. As a result, increasing the speed leads the extension of the
chains along the fiber direction, and therefore an increase in the
intercrystallite distance. For reeling speeds beyond $10\,\mathrm{mm/s}$,
the fast pulling motivates the merging of the chains, which is accompanied
by a reduction in the diameter of the silk. In parallel, intra-molecular
$\beta$-sheets form \citep{Du2011}, and these packed structures
along the macro-chains decrease the intercrystallite distance. 

Next, we determine the orientation parameter $\odfpar$. To this end,
we employ the measurements of the orientation function in the work
of \citet{Du2006} along with Eqs. \ref{eq:orientation f} and \ref{eq:cos_sq}.
It is convenient to fit the orientation parameter 
\begin{equation}
\odfpar\left(\reelingsp\right)=\bar{\odfpar}\left(1-\exp\left(-\tau_{\odfpar}\,\reelingsp\right)\right),\label{eq:b_eq}
\end{equation}
where $\bar{\odfpar}=25.5$ is the value at $\reelingsp\rightarrow\infty$
and $\tau_{\odfpar}=0.045$ is a rate constant. Eq. \ref{eq:b_eq}
captures the experimental results, as shown in Fig. \ref{fig:du_exp_model_parameters}c.

To characterize the amorphous domain, we examine the ratio $\ratref$.
As previously stated, increasing the reeling speed aligns and stretches
the macro-chains, thereby increasing $\macchainref$. In parallel,
higher reeling speeds lead to an increase in intra-molecular $\beta$-sheets
and a decrease in crystallinity \citep{Du2011}. These are expected
to affect the number of repeat units $\links$ along the macro-chain.
Therefore, the trend of $\ratref$ is not clear and requires further
experimental investigation. For simplicity, in this work we propose
the expression 
\begin{equation}
\ratref\left(\reelingsp\right)=\bar{\ratref}\left(1-\exp\left(-\tau_{\ratref}\,\reelingsp\right)\right),\label{eq:rat_ref_exp}
\end{equation}
which assumes an initial increase in $\ratref$ up to a limiting value
$\bar{\ratref}$ at $\reelingsp\rightarrow\infty$ with a rate constant
$\tau_{\ratref}$. Once the ratio $\ratref$ is obtained, $\langref$
is determined from the Langevin function using the approximation of
\citet{Cohen1991}. The number of repeat units $\links\left(\reelingsp\right)=\macchainref\left(\reelingsp\right)/\linklen\ratref\left(\reelingsp\right)$,
where Eqs. \ref{eq:R_eq} and \ref{eq:rat_ref_exp} are employed.
For simplicity, we assume that the number of intermolecular bonds
$\bonds=13$ is independent of the reeling speed. 

The parameter $\shear$ is proportional to the number of representative
elements in the fiber. At faster reeling speed, the crystallite size
and the diameter of the fiber decrease. Moreover, the increase in
alignment leads to an increase in nucleation of crystalline domains
\citep{Du2006}, and as such an increase in the number of representative
elements. Therefore, one can expect $\shear$ to increase monotonically
up to a limiting value. To capture this effect, we propose the expression
\begin{equation}
\shear\left(\reelingsp\right)=\bar{\shear}\left(1-\exp\left(-\tau_{\shear}\,\reelingsp\right)\right),\label{eq:shear_exp}
\end{equation}
where $\bar{\shear}$ denote the limiting value at $\reelingsp\rightarrow\infty$
and $\tau_{\shear}$ is a rate constant. 

\begin{figure}
\begin{centering}
\includegraphics[width=6cm]{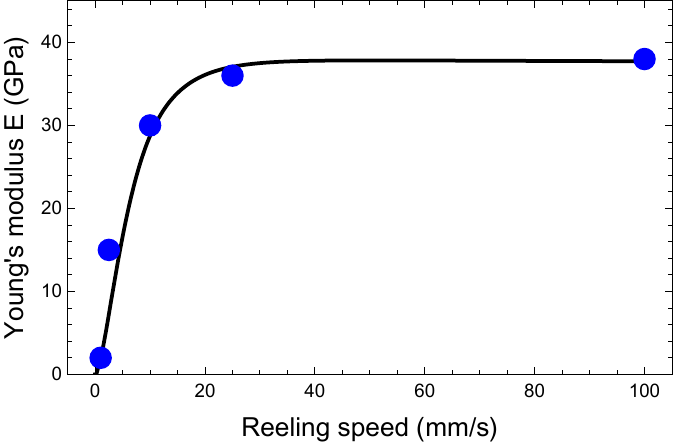}
\par\end{centering}
\caption{The Young's modulus $E$ (Eq. \ref{eq:Young_modulus}) as a function
of the reeling speed $\protect\reelingsp$. The circle marks denote
the experimental results of \citet{Du2006}. \label{fig:du_exp_fit}}
\end{figure}

To determine the Young's modulus, we first compute $\ratref$ and
$\shear$. To this end, we fit $\bar{\ratref}=0.9$, $\tau_{\ratref}=10$,
$\bar{\shear}=6.8\,\mathrm{MPa}$, and $\tau_{\shear}=0.15$ to Eqs.
\ref{eq:rat_ref_exp} and \ref{eq:shear_exp}. Fig. \ref{fig:du_exp_fit}
depicts the Young's modulus $E$ (Eq. \ref{eq:Young_modulus}) as
a function of the reeling speed $\reelingsp$, and we find that the
model is capable of capturing the experimental findings. 

\section{The influence of microstructure on stiffness }

To better understand the influence of microstructure on stiffness
and demonstrate the merit of the proposed framework in the design
of bio-mimetic fibers, in this section we investigate the influence
of the initial chain stretch, the number of intermolecular bonds,
the crystallinity, and the chain alignment on stiffness. To this end,
we consider the normalized stiffness $E/\left(3\ues\boltzmann T\right)$,
where the normalization factor $3\ues\boltzmann T$ is the stiffness
of a network with long chains characterized by the average end-to-end
vector $\macchainref=\sqrt{\links}\linklen$ such that $\ratref=1/\sqrt{\links}$
\citep{Treloar1975,Flory1953}. 

\subsection{The effect of fiber composition on stiffness}

We begin by considering fibers with randomly oriented and uniformly
distributed representative elements (i.e. $\odfpar\rightarrow0$).
We examine three fiber microstructures: (1) a rubbery polypeptide
network with no intermolecular bonds and no crystalline domains ($\bonds=0$
and $\cref=0$), (2) a network of interacting macro-chains with $\bonds$
intermolecular bonds and no crystalline domains ($\cref=0$), and
(3) a network with a volume fraction $\cdfrac$ of crystalline domains
and no intermolecular bonds $\bonds=0$. 

\begin{figure}
\hspace*{\fill}(a) \includegraphics[width=4.5cm]{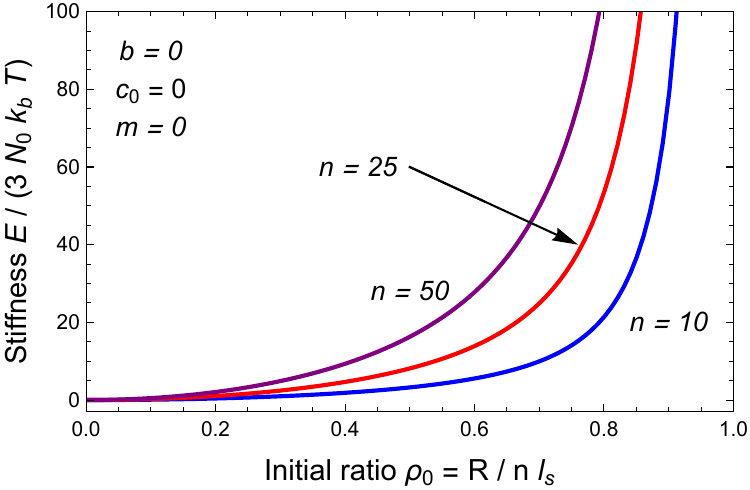}~~(b)
\includegraphics[width=4.5cm]{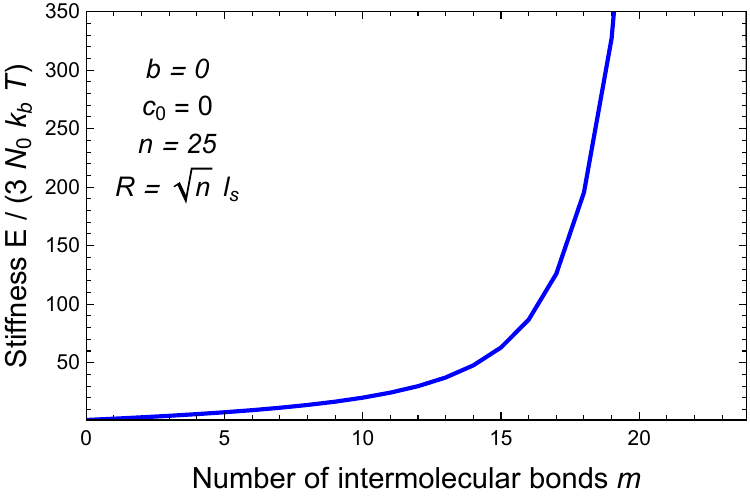}~~(c) \includegraphics[width=4.5cm]{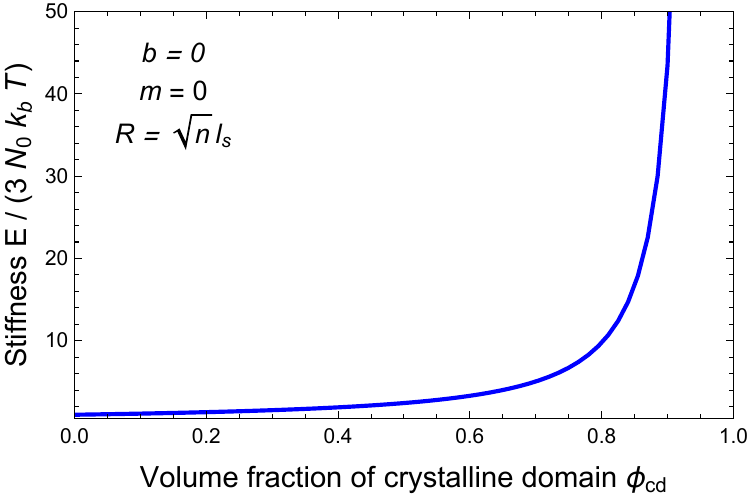}\hspace*{\fill}

\caption{The normalized stiffness $E/\left(3\protect\ues\protect\boltzmann T\right)$
as a function of (a) the initial ratio $\protect\macchainref/\protect\links\protect\linklen$
in a rubbery network with $\protect\bonds=0$ and $\protect\cref=0$,
(b) the number of intermolecular bonds $\protect\bonds$ in a fiber
with no crystalline domains ($\protect\cref=0$), and (c) the volume
fraction $\protect\cdfrac$ of the crystalline domain with $\protect\bonds=0$.
\label{fig:stiffness_structure_b0}}
\end{figure}

Fig. \ref{fig:stiffness_structure_b0}a explores the influence of
the pre-stretch of the polypeptide chains by depicting the normalized
stiffness $E/\left(3\ues\boltzmann T\right)$ as a function of the
initial ratio $\ratref=\macchainref/\links\,\linklen$ of macro-chains
with $\links=10,25,$ and $50$ repeat units in a rubbery network
with $\bonds=0$ and $\cref=0$. In the case of $\ratref\rightarrow0$
the chains are extremely soft and the network has practically zero
stiffness. Freely jointed chains stiffen with pre-stretch, and therefore
as the ratio $\ratref$ increases the modulus of the network increases.
Interestingly, for a given ratio $\ratref$ we see that decreasing
the number of repeat units leads to a softening of the fiber. This
result may seem somewhat counterintuitive, since typically shorter
chains are stiffer. However, the initial end-to-end distance $\macchainref=\ratref\links\linklen$
increases with $\links$ and the non-linear response of the chains
explains this result. This shows that increasing the initial extension
of the chains leads to stiffer fibers, as expected.

Fig. \ref{fig:stiffness_structure_b0}b depicts the normalized stiffness
$E/\left(3\ues\boltzmann T\right)$ as a function of the number of
intermolecular bonds $\bonds$. The macro-chains assume the average
end-to-end distance $\macchainref=\sqrt{\links}\linklen$ and we set
$\links=25$ and $\bstiffness=10$. Increasing the number of intermolecular
bonds leads to: (1) the addition of load carrying agents and (2) the
shortening of the freely jointed chains along the representative elements.
Owing to these two factors, the fiber stiffens with increasing $\bonds$. 

The effect of crystallinity is shown in Fig. \ref{fig:stiffness_structure_b0}c,
which plots the normalized stiffness $E/\left(3\ues\boltzmann T\right)$
as a function of the volume fraction of crystalline domains $\cdfrac=\cref/\lueref$.
As in the previous case, we set the end-to-end distance $\macchainref=\sqrt{\links}\linklen$,
the repeat unit length $\linklen=0.05\lueref$, the number of repeat
units $\links=\left(\lueref\left(1-\cdfrac\right)/\linklen\right)^{2}$
(Eq. \ref{eq:ref_len}), and $\cdstiffness=100$. Since the chains
in the network are randomly oriented and uniformly distributed, $\cdfrac=0$
leads to $E/\left(3\ues\boltzmann T\right)=1$. As $\cdfrac$ increases,
the fiber stiffens non-linearly as expected due to two main factors:
(1) the crystalline domains give rise to high localized stiffness
and (2) increase in crystallinity leads to shorter polypeptide chains,
which tend to be stiffer. Therefore, controlling the combination of
these two factors allows one to enhance the stiffness of fibers. 

\subsection{The effect of chain alignment on stiffness}

Next, we investigate the influence of alignment on the overall response
of bio-fibers. To this end, we follow the previous section and study
three fiber microstructures: (1) a rubbery polypeptide network with
no intermolecular bonds and no crystalline domains ($\bonds=0$ and
$\cref=0$), (2) a network of interacting macro-chains with $\bonds$
intermolecular bonds and no crystalline domains ($\cref=0$), and
(3) a network with a volume fraction $\cdfrac$ of crystalline domains
and no intermolecular bonds $\bonds=0$. The following simulations
assume that the initial end-to-end distance of the chains is $\macchainref=\sqrt{\links}\linklen$
such that $\ratref=1/\sqrt{\links}$. 

\begin{figure}
\hspace*{\fill}(a) \includegraphics[width=4.5cm]{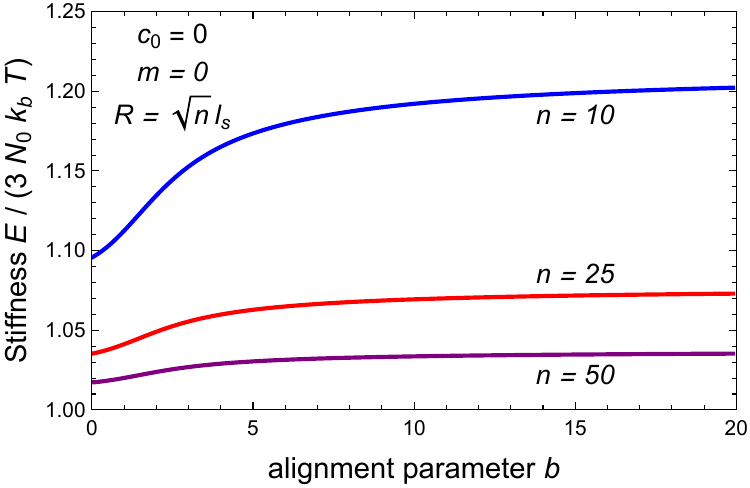}~~(b)
\includegraphics[width=4.5cm]{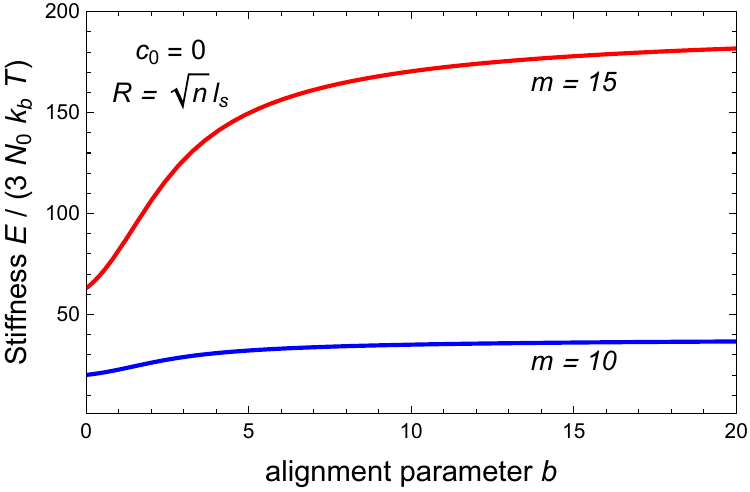}~~(c) \includegraphics[width=4.5cm]{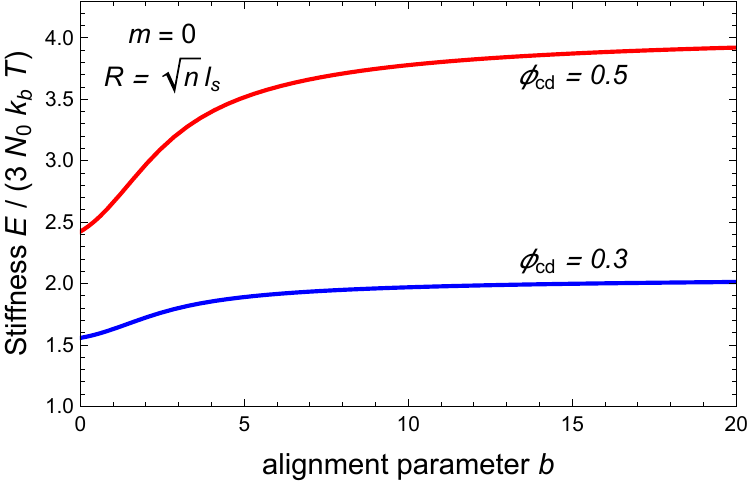}\hspace*{\fill}

\caption{The normalized stiffness $E/\left(3\protect\ues\protect\boltzmann T\right)$
as a function of the alignment parameter $\protect\odfpar$ for networks
characterized by (a) $\protect\bonds=0$ and $\protect\cref=0$, (b)
$\protect\cref=0$ and $\protect\bonds=10,15$, and (c) $\protect\bonds=0$
with $\protect\cdfrac=0.3,0.5$. \label{fig:stiffness_structure_b}}
\end{figure}

Fig. \ref{fig:stiffness_structure_b}a investigates the influence
of alignment on different chain lengths by depicting the normalized
stiffness $E/\left(3\ues\boltzmann T\right)$ as a function of $\odfpar$
for a fiber with macro-chains with $\links=10,25,$ and $50$ repeat
units in a rubbery network with $\bonds=0$ and $\cref=0$. We find
that shorter chains exhibit a higher modulus, which is explained by
the higher ratio $\ratref=1/\sqrt{\links}$ for lower $\links$. Recall
that as as $\ratref$ increases, the chains stiffen locally and therefore
the overall modulus is higher. As expected, as the parameter $\odfpar$
increases, accounting for a higher degree of chain alignment, a stiffening
is observed. 

Fig. \ref{fig:stiffness_structure_b}b explores the influence of intermolecular
bonds and chain alignment on the overall response by plotting $E/\left(3\ues\boltzmann T\right)$
as a function of the alignment parameter $\odfpar$ for a network
with $\bonds=10$ and $\bonds=15$ intermolecular bonds, macro-chains
with $\links=25$, bond stiffness $\bstiffness=10$, and no crystalline
domains ($\cref=0$). Stronger chain alignment forces the bonds to
form perpendicular to the fiber axis and leads to an increase in the
modulus. In the general case, an increase in the alignment parameter
leads to a higher density of intermolecular bonds since more load-carrying
constituents are required to counteract the entropic forces and allow
for the anisotropic distribution of the chains in the network. This
leads to further stiffening since the increase in intermolecular bonds
results in shorter freely jointed chains, which is another factor
that is conducive to the stiffening of the fiber. 

Fig. \ref{fig:stiffness_structure_b}c studies the influence of the
alignment parameter $\odfpar$ on the normalized stiffness $E/\left(3\ues\boltzmann T\right)$
for two degrees of crystallinity, characterized by the volume fractions
$\cdfrac=0.3$ and $0.5$. The fiber has no intermolecular bonds ($\bonds=0$),
the repeat unit length $\linklen=0.05\lueref$, the number of repeat
units $\links=\left(\lueref\left(1-\cdfrac\right)/\linklen\right)^{2}$
(Eq. \ref{eq:ref_len}), and $\cdstiffness=100$. As expected, a higher
degree of alignment stiffens the fiber. For a given microstructural
alignment $\odfpar$, we find that fibers with higher crystallinity
are stiffer. This stems from the contribution of additional stiff
crystalline domains in the network. 

The above study shows that an increase in the stiffness of the fiber
can be achieved in several ways: (1) increase the pre-stretch of the
chains in the initial undeformed state, (2) increase in the density
of intermolecular bonds, (3) increase in the volume fraction of crystalline
domains, and (4) enhance the alignment of the chains in the network. 

\section{Influence of temperature on stiffness }

\global\long\def\Tone{T_{1}}%
\global\long\def\Ttwo{T_{2}}%
\global\long\def\Td{T_{d}}%
\global\long\def\Trt{T_{RT}}%

Recent works investigated the dependence of the Young's modulus of
bio-fibers on temperature \citep{Guan2013,Mazzi2014,Yang2005}. These
works typically begin with a fiber at a given temperature, and measure
the stiffness as the temperature increases. With the aim of controlling
the stiffness at various temperatures, we explore the thermally-induced
microstructural evolution and its influence on stiffness. To this
end, it is convenient to distinguish between three temperature ranges:
(1) a temperature $\Trt\leq T<100\text{�}$, where $\Trt$ is the
room temperature, (2) $100\text{�}\leq T<\Td$, where $\Td>200\text{�}$
is the temperature at which the polypeptide chains and the crystalline
domains begin to degrade, and (3) $\Td\leq T$. 

\begin{figure}
\begin{centering}
\includegraphics[width=16cm]{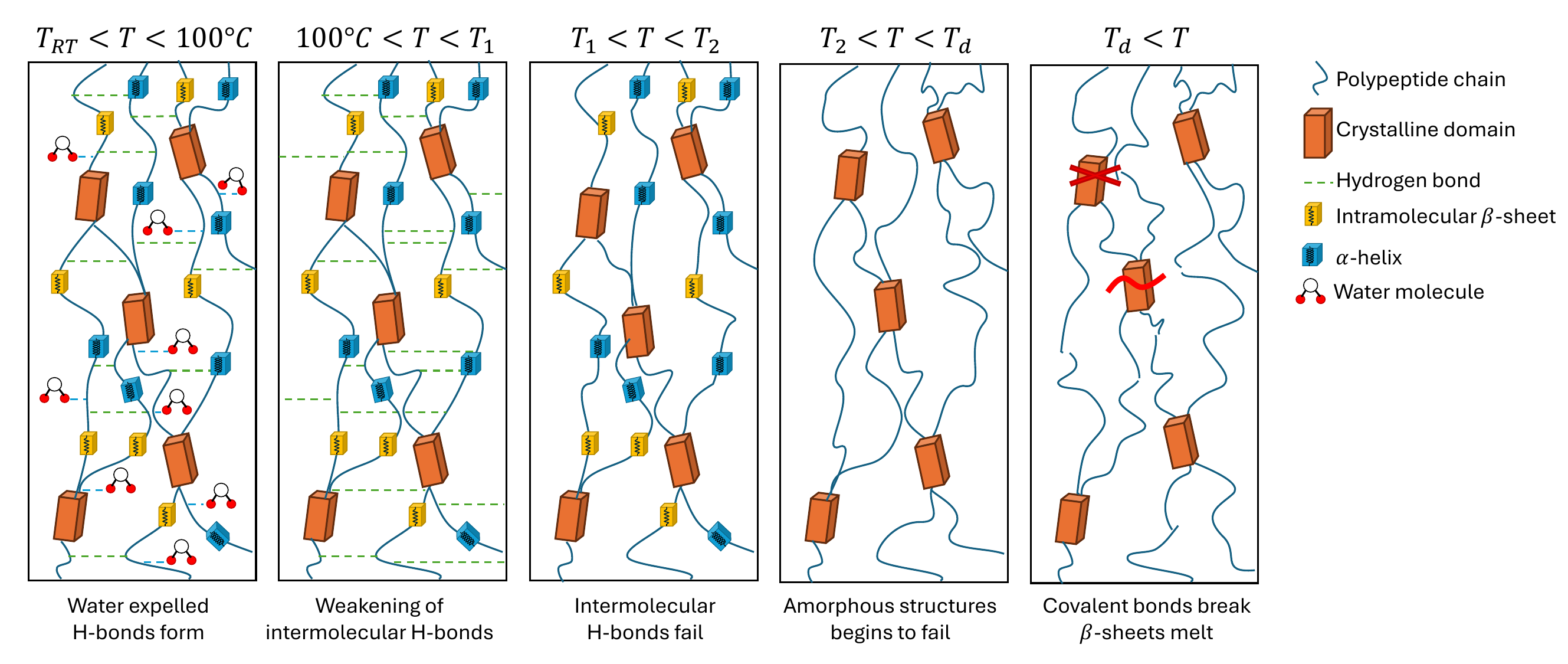}
\par\end{centering}
\caption{The microstructural evolution of a bio-fiber as the temperature increases.
\label{fig:microstructure_temp}}
\end{figure}

In the first range, characterized by $\Trt<T<100\text{�}$, as the
temperature increases water is expelled from the fiber, as demonstrated
in Fig. \ref{fig:microstructure_temp}. The crystalline domains are
not affected and the change in the intercrystallite distance is minor,
and therefore we assume that $\normL$ is constant. Furthermore, the
outflux of water leads to a slight compaction of the fiber and a minor
disordering of the crystalline domains along the $\beta$-sheet directions
\citep{Guan2013}. We assume that these effects are negligible and
set the parameters $\shear$, $\odfpar$, and $\ratref$ as constants.

We follow by examining the intermolecular bonds. Recall that water
molecules hinder the formation of intermolecular interactions by forming
hydrogen bonds with the peptide chains. Therefore, as water diffuses
out of the fiber the number of intermolecular bonds $\bonds$ increases
\citep{Ozeren2021,Greco2025,Mazzi2014}. In addition, as the temperature
increases the intrinsic energy of the bonds changes. Specifically,
heating the fiber adds kinetic energy, resulting in molecular vibrations
that weaken the stability of the intermolecular hydrogen bonds \citep{Hu2025}.
This results in a decrease in the bond stiffness $\bstiffness$ . 

Overall, in the range $\Trt<T<100\text{�}$, the water outflux leads
to the addition of the intermolecular hydrogen bonds and an overall
increase in stiffness with temperature \citep{Guan2013,Greco2025,Yang2005,Latza2015}. 

In the second range of temperatures $100\text{�}\leq T<\Td$, illustrated
in Fig. \ref{fig:microstructure_temp}, the crystalline domains remain
largely unaffected \citep{Guan2013}. However, experiments show that
temperature increase leads to two main microstructural changes within
the amorphous zone, which occur at temperatures $\Tone$ and $\Ttwo>\Tone$
between $100\text{�}$ and the degradation temperature $\Td$ \citep{Yuan2010,Guan2013}.
In the range $100\text{�}\leq T<\Tone$, the intermolecular interactions
continue to weaken, while the overall microstructural changes are
minor. Once the temperature rises above $\Tone$, the intermolecular
bonds begin to dissociate. As $\bonds$ decreases, freely jointed
chains gain mobility as the number of repeat units $\links/\left(\bonds+1\right)$
increases. Note that the additional entropic freedom is expected to
result in a slight decrease in the intercrystallite distance $\macchainref$,
and consequently $\normL$. In addition, the chains can reorient to
occupy a more uniformly distributed distribution, characterized by
a decrease in the parameter $\odfpar$. As the temperature further
increases beyond $\Ttwo$, secondary structures along the polypeptide
chain begin to break down through the cooperative dissociation of
intramolecular hydrogen bonds \citep{Qi2017,Guan2013,Mazzi2014,Du2011,Roemer2008,Yuan2010}.
This leads to an overall increase in the number of repeat units $\links$
along a macro-chain, which results in additional entropic mobility,
and an increase in the intercrystallite distance $\macchainref$. 

It is important to note that the temperatures $\Tone$ and $\Ttwo$
have a wide range, which depends on many factors, including the type
of fiber, the species, and the retrieval process \citep{Guan2013,Dao2017,Yuan2010}.

The third temperature range $T>\Td$ is characterized by the degradation
of the proteins in the fiber. Specifically, the crystalline domains
and the covalent bonds along the polypeptide chains can fail \citep{Guan2013,Yuan2010,Cebe2013}.
As a result, the fiber loses its structural integrity and a significant
softening is observed. Typically, $\Td\sim250-350\text{�}$.

With the aim of providing a quantitative method that enables one to
evaluate the microstructural evolution of the fiber, we employ the
experimental findings of \citet{Guan2013} which measured the stiffness
as a function of temperature of \emph{Bombyx mori Silkworm Silk}.
For simplicity, we assume that the orientational changes and the increase
in the intercrystallite distances are minor and the parameter $\shear$,
which accounts for the number of representative elements, is constant
throughout. In accordance with Table \ref{tab:E_comparison_properties},
we set $\normL=17.7$ and $\odfpar=13.7$. In the work of \citet{Guan2013},
the Young's modulus in room temperature is lower than the average
for cocoon silk ($\sim10\,\mathrm{GPa}$, as shown in Table \ref{tab:E_comparison_properties}).
Accordingly, we set $\mu=21\,\mathrm{MPa}$. 

The temperatures $\Tone=165\text{�}$ marks the dissociation of the
intermolecular hydrogen bonds and $\Ttwo=190\text{�}$ denotes the
initiation of the breaking of secondary intramolecular structures
\citep{Yuan2010,Guan2013}. We underscore that further research into
the influence of temperature on these key quantities is required. 

As described by \citet{Hu2025}, the stiffness of the intermolecular
bonds decreases with temperature, and we propose the expression 
\begin{equation}
\bstiffness\left(\temp\right)=\tilde{\bstiffness}-\tau_{\stiffness}\ln\left(T-T_{RT}\right),\label{eq:bond_stiffness_temp}
\end{equation}
where $\tilde{\bstiffness}=10$ is the stiffness of the hydrogen bonds
at room temperature and $\tau_{\stiffness}=0.1$ is a rate constant. 

To capture the dissociation and the formation of intermolecular hydrogen
bonds, we employ the piece-wise function
\begin{equation}
\bonds\left(\temp\right)=\begin{cases}
\bonds_{0}-\tau_{\bonds}\left(100-T\right) & \Trt\leq\temp<100\text{�}\\
\bonds_{0} & 100\text{�}\leq T<\Tone\\
\bonds_{0}\exp\left(-\tau_{g}\left(\temp-\Tone\right)\right) & \Tone<\temp<\Td
\end{cases}.\label{eq:hbond_increase}
\end{equation}
Here, $\bonds_{0}=10$ is the number of intermolecular bonds at $T=100\text{�}$
and $\tau_{\bonds}=0.053$ and $\tau_{g}=0.009$ are rate constants. 

Lastly, the number of repeat units 
\begin{equation}
\links\left(\temp\right)=\begin{cases}
\links_{0} & \Trt\leq\temp<\Ttwo\\
\links_{0}\exp\left(\tau_{n}\left(T-\Ttwo\right)\right) & \Ttwo\leq\temp<\Td
\end{cases},\label{eq:links_increase}
\end{equation}
where $\links_{0}=12$ (as in Table \ref{tab:E_comparison_properties})
and $\tau_{n}=2.3\cdot10^{-3}$. Following the assumption that the
changes in the intercrystallite distance are relatively small, we
set the ratio $\ratref=\macchainref/\linklen\,\links\left(T\right)$,
where $\macchainref=8.7\,\mathrm{nm}$ is the intercrystallite distance. 

\begin{figure}
\begin{centering}
\includegraphics[width=7cm]{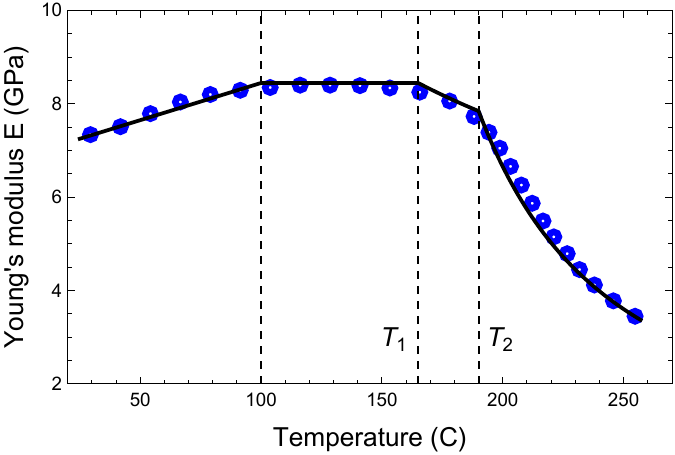}
\par\end{centering}
\caption{The Young's modulus $E$ (Eq. \ref{eq:Young_modulus}) as a function
of the temperature $T$. The continuous curve correspond to the model
predictions (Eq. \ref{eq:Young_modulus}) and the circle marks denote
the experimental results of \citet{Guan2013}. \label{fig:E_temp}}
\end{figure}

Fig. \ref{fig:E_temp} depicts the Young's modulus $E$ (Eq. \ref{eq:Young_modulus})
as a function of the temperature $T$, where Eqs. \ref{eq:bond_stiffness_temp},
\ref{eq:hbond_increase}, and \ref{eq:links_increase} are employed.
The continuous curve correspond to the model predictions (Eq. \ref{eq:Young_modulus})
and the circle marks denote the experimental results of \citet{Guan2013}.
The proposed model is capable of capturing the experimental findings
while capturing the underlying microstructural evolution with temperature. 

\section{Conclusions}

In this work, an energetically motivated microstructural framework
was developed to provide a fundamental, quantitative understanding
of the underlying physical mechanisms governing the stiffness of biological
fibers. The derived model explicitly accounts for three key concurrent
deformation mechanisms: (1) the entropic extension of polypeptide
chains within the amorphous matrix, (2) the elastic stretching and
cooperative rotation of rigid crystalline domains, and (3) the relative
sliding of chains enabled by the distortion and ultimate dissociation
of weak intermolecular interactions. By employing tools from statistical
mechanics to capture the non-linear entropic response of the polypeptide
chains and proposing simple expressions for the behavior of crystalline
domains and intermolecular bonds, a closed-form analytical expression
for the Young's modulus was obtained. We underscore that the model
parameters have a well-founded physical basis, and we outline how
key model inputs can be systematically approximated through experimental
and numerical characterization techniques. 

The predictive capabilities and physical merit of the developed framework
were demonstrated through comparison to various experimental findings
on different types of bio-fibers. Specifically, the model successfully
captures the characteristic stiffness of spider silk and cocoon silk.
Furthermore, the model was applied to understand the relationship
between reeling speed, microstructural reorganization, and fiber stiffness
in spider silk. The ability of the model to capture the stiffness
of fibers retrieved via different reeling speeds strengthens its merit
and highlights its potential in translating microstructural variations
into macroscopic properties. 

To better understand the influence of microstructure on performance
and provide a pathway to the design of bio-mimetic fibers, we employed
the model to study the relationship connecting key microstructural
parameters---such as polypeptide chain pre-stretch, the density of
intermolecular bonds, crystallinity, and network alignment---to the
overall fiber stiffness. We demonstrated that increasing these structural
quantities leads to a stiffer fiber, and thus argue that the developed
framework provides a robust tool to systematically quantify these
enhancements. 

Lastly, we investigated the effects of temperature on the microstructure
of bio-fibers. To this end, we distinguish between three ranges of
temperature: (1) $\Trt\leq T<100\text{�}$, (2) $100\text{�}\leq T<\Td$,
and (3) $\Td\leq T$. Here, recall that $\Td>200\text{�}$ is the
temperature at which the proteins and the crystalline domains begin
to degrade. In the first range, the stiffness increases mostly due
to an outflux of water molecules and the formation of additional intermolecular
hydrogen bonds. In the second range of temperatures, the modulus decreases
as a result of structural changes within the amorphous zone that provide
additional chain mobility, such as the thermally-induced weakening
and dissociation of intermolecular and intramolecular bonds. Additional
microstructural changes such as an increase in the intercrystallite
distance, loss of alignment, and the dissociation of weak intra- and
inter-molecular bonds may also occur and require further investigation.
For temperatures above $\Td$ the proteins degrade and the fiber loses
its structural integrity, resulting in a drastic loss of stiffness.
To demonstrate the merit of the proposed framework, we fit the model
to experimental data of the stiffness as a function of temperature
in \emph{Bombyx mori Silkworm Silk}. The model is capable of capturing
the overall trends and sheds light on the microstructural evolution. 

In conclusion, the insights and the quantitative relationships developed
in this work shed light on the interplay between the microstructure
and the stiffness of biological fibers. We also show how processing
parameters and temperature influence the microstructure and, consequently,
the stiffness. Ultimately, this framework provides the fundamental
design principles required to guide, tune, and optimize, the design
of bio-inspired fibers with tailored mechanical performance. 

\bibliographystyle{biochem}
\bibliography{refs}

\end{document}